\documentclass[12pt,a4]{article}

\usepackage{amsthm,amsmath,latexsym,amssymb,amsfonts,amscd}
\usepackage{graphics,lscape,fancyhdr,array,stmaryrd,euscript}
\usepackage{empheq}
\usepackage{dsfont}
\usepackage{verbatim}
\usepackage{color,tikz,tikz-cd}
\usetikzlibrary[snakes]
\usepackage{relsize,slashed}
\numberwithin{equation}{section}
\usepackage{hyperref}

\newcommand{\bep}{\begin{picture}}
\newcommand{\eep}{\end{picture}}

\newcounter{YoungHeight}\newcounter{YoungWidth}

\newcounter{Mul1}\newcounter{Mul2}\newcounter{Mul3}\newcounter{Mul4}
\newcounter{A0}\newcounter{A1}\newcounter{A2}
\newcounter{B3}
\newcounter{C3}\newcounter{C4}
\newcounter{D1}\newcounter{D2}\newcounter{D3}
\newcounter{T0}\newcounter{T1}

\newlength{\txtHShift}

\newlength{\txtWidth}

\newcommand{\Add}[3]{\setcounter{#1}{#2}\addtocounter{#1}{#3}}

\newcommand{\Length}[1]{#10}
\newcommand{\YoungScale}{}

\newcommand{\BlockApar}[2]{\parbox{\Length{#1}pt}{\YoungScale\bep(\Length{#1},\Length{#2}){\Add{A1}{#1}{1}\Add{A2}{#2}{1}}%
\multiput(0,0)(10,0){\value{A1}}{\line(0,1){\Length{#2}}}\multiput(0,0)(0,10){\value{A2}}{\line(1,0){\Length{#1}}}%
\setcounter{YoungHeight}{\Length{#2}}\setcounter{YoungWidth}{\Length{#1}}\eep}}

\newcommand{\YoungpA}{\BlockApar{1}{1}}
\newcommand{\YoungpB}{\BlockApar{2}{1}}

\newcommand{\YoungpAA}{\BlockApar{1}{2}}

\usepackage{setspace}

 \usepackage[numbers,sort&compress]{natbib}
 \usepackage[nottoc]{tocbibind}

\newcommand{\besubeqs}{\begin{subequations}}
\newcommand{\esubeqs}{\end{subequations}}

\newcommand{\fdu}[2]{{}_{#1}{}^{#2}\,}

\newcommand{\hs}{\mathfrak{hs}}

\begin{document}
\pagenumbering{gobble}
\hfill
\vspace{-1.5cm}
\vskip 0.05\textheight
\begin{center}
{\Large\bfseries Conformal Galilean Spin-3 Gravity in 3d
}

\vspace{0.4cm}
\vskip 0.03\textheight
I. \textsc{Lovrekovic}$^*$, 
K. \textsc{Schaefer}$^*$

\vskip 0.03\textheight

\vspace{5pt}

{\it $^*$ Institute for Theoretical Physics, Technische Universit\"at Wien, \\ Wiedner Hauptstrasse, 8-10,\\
1040, Wien, Austria}

\end{center}

\vskip 0.02\textheight

\begin{abstract}
We consider Galilean limit of conformal algebra for spin-2 and spin-3 fields, and study the gauge theory of these algebras. We analyze the equations of motion and obtain the spectra of the theories. 
\end{abstract}

\pagenumbering{arabic}
\setcounter{page}{1}

\section{Introduction}


Galilean theories have been studied in many works in different number of dimensions  \cite{Bergshoeff:2022eog}
\cite{Bergshoeff:2022iyb}
\cite{Bergshoeff:2022qkx}
\cite{Bergshoeff:2020fiz}
\cite{Bergshoeff:2019ctr}
\cite{Bergshoeff:2019vlk}. 
When we say Galilean theory we have to specify it in the context of gauge/gravity duality. Number of studies were devoted to Galilean (conformal) field theories \cite{Bagchi:2009my}
\cite{Bagchi:2009ca}
\cite{Bagchi:2010vw}
\cite{Bagchi:2014ysa}
\cite{Bagchi:2013toa}, which, together with similar non-relativistic theories, were found to be manifested in the condensed matter systems. In this work, we will focus on the gravity side.  

In general, Galilean theory is invariant under the Galilean transformations. In the non-conformal case Galilean symmetries arise after In\"on\"u Wigner contraction of the Poincare symmetries. 
 This picture can be viewed as a light cone completely opened, where speed of light goes to infinity, i.e. the light rays of a light cone are closing 0 and 180 degrees angles. One obtains the space where exchange of information is possible with each point at any time. Galilean symmetries hugely depend on the number of dimensions in which we are considering them. In two dimensions they had additional application because they can be interchanged with Carrolian symmetries by exchanging the time and space component. 

Galilean theory has been studied in four dimensions in the relation with Newton Cartan theory. Where it was shown that one cannot obtain Newton-Cartan gravity by gauging the Galilean algebra, instead, one has to add a central extension to it, obtaining Bargmann algebra. Gauging Bargmann algebra  reproduces Newton-Cartan gravity. The metric of the Galilean gravity is degenerate and one can't solve the equations for all components of the spin connection. This is solved by the central extension, in Bargmann algebra. 
Importance of the Newton-Cartan gravity is in our current understanding of the universe at the non-relativistic scales.  Newton-Cartan theory is obtained in the non-relativistic limit of Einstein gravity. 
The study in \cite{Banerjee:2016laq} showed that one can obtain torsional Newton-Cartan geometry from
Galilean gauge theory in 4d. 
\\
Interesting suggestion, however much less studied in this context, included addition of conformal invariance \cite{Andringa:2010it}. The analysis of conformal Galielean theories have so far been focused on electrodynamics \cite{Bagchi:2014ysa} and boundary theories. Boundary theories are here referred to in the context of AdS/CFT duality, where the conformal field theory at the boundary has symmetries of non-relativistic groups, among which are also Galilean symmetries 
\cite{Nishida:2007pj,Taylor:2008tg,Taylor:2015glc}. 
\\
\indent Conformal symmetry was shown to be very important in physics. 't Hooft has speculated that conformal symmetries plays a key role at the physics at the Planck energy scale \cite{tHooft:2015vaz}. He worked on a number of studies analyzing the conformal generalizations of Einstein gravity \cite{tHooft:2010mvw,tHooft:2010xlr}. In four dimensions, it was shown that a theory with conformal symmetry can explain galactic rotation curves without the addition of dark matter \cite{Mannheim:1988dj}, and it can be considered as a well defined theory \cite{Grumiller:2013mxa,Irakleidou:2014vla}.  The negative sides are that theory has ghosts which are usually treated with Pais-Uhlenbeck \cite{Pais:1950za} oscillator, however purely on theoretical side, scientists often study it as a toy model for its additional symmetry.  In two dimensions the symmetry plays important role in the field theories dual to higher dimensional bulk gravity theories, in AdS/CFT duality. In three dimensions conformal gravity theory can be written as a gauge theory of conformal algebra in terms of Chern-Simons action \cite{Horne:1988jf}.

 Higher spin gravities of In\"on\"u Wigner contracted algebras have been considered in \cite{Bergshoeff:2016soe,Lovrekovic:2021dvi}. \cite{Bergshoeff:2016soe} studied gravity theories of spin three fields in non- and ultra-relativistic limit, while  \cite{Lovrekovic:2021dvi} focused on the spectrum and construction of the conformal gravity theory in the ultra-relativistic limit. The construction of the algebras for the higher spin gravities in these limits was recently described in \cite{Campoleoni:2021blr} using similar methods as in \cite{Ammon:2020fxs}. 

Here, we study Galilean conformal gravity and its spin-3 generalization in three dimensions, obtaining it as an In\"on\"u Wigner contraction of the conformal algebra, and conformal spin-3 algebra respectively. This leads to the spectrum of the fields in the theory, equations of motion, and constraints. Algebra of conformal graviton and conformal spin-3 field in our construction is isomorphic to the $sl(5)$ algebra, so one way to look at the contraction is as Galilean limit of $sl(5)$ algebra. 
The paper is structured as follows. In the second section we overview Chern-Simons theory as a gauge theory of $so(3,2)$ group, and the construction of the algebras that we use as gauge algebras. We also outline the contraction of algebras previously constructed. In section three we consider Galilean conformal gravity as a gauge theory of Galilean conformal algebra (GCA), while in section four we consider spin three Galilean conformal gravity.

\section{Chern-Simons theory}

Our starting point is the three dimensional Chern-Simons action of the form 
\begin{align}
S=\int \langle A\wedge dA +\frac{2}{3}A\wedge A\wedge A \rangle.\end{align}
 A is Lie algebra valued one form, and depending on the algebra in which we value the field A, one obtains theories of the different properties. For the underlying gauge algebra  $sl(2,\mathbb{R})\oplus sl(2,\mathbb{R})$ the theory corresponds to three dimensional Einstein gravity, and 
 it can be generalized to higher spin algebra in which case one obtains the theory for massless higher spins in 3d \cite{Blencowe:1988gj}. When the underlying algebra is conformal one obtains three dimensional conformal gravity \cite{Horne:1988jf}, which can also be generalized to higher spin theory \cite{Pope:1989sr}, \cite{Grigoriev:2019xmp}. 
The specific property of the theory in \cite{Grigoriev:2019xmp} is that it contains infinite number of theories with finite number of fields of arbitrary spin. To take the non-relativistic limit of this theory we first take the limit of the constructed underlying algebras, and then evaluate in them the field A and corresponding gauge parameter. This leads to the equations of motion that we can solve to obtain the spectrum of the theory. 


\subsection{Construction}


To consider the non-relativistic limit of GCA and its generalization to arbitrary higher spin case, first
we need to know the $so(3,2)$ higher spin extension, and the algebras that have its nontrivial representation with $so(3,2)$ as a subalgebra.
 By taking nontrivial finite-dimensional irreducible representation $V$ of the $so(3,2)$ one can construct finite spectrum of higher spin fields. $V$ is irreducible tensor or a spin-tensor, evaluated in the $U(so(d,2))$. To evaluate $V$ one multiplies the $so(3,2)$ tensors $T_{AB}=-T_{BA}$ in the representation and determines the generated algebra  \cite{Grigoriev:2019xmp}. Here, we will denote the indices of $so(3,2)$ with $A,B,..=0,..,4$ and an invariant metric with $\eta_{AB}$. 

Generally this algebra is denoted with $\hs(V)$. 
To see how one obtains spin-2 and spin-3 algebra in this decomposition we take
 the vector representation,  denoted by  Young diagram $\YoungpA$ that has one-cell. The corresponding algebra has the spectrum, $\hs(\YoungpA)$:
\begin{align}
    \mathfrak{hs}(\YoungpA)=&\YoungpA\otimes \YoungpA= \YoungpAA \oplus \YoungpB\oplus \bullet .
\end{align}
Algebra that we consider is a matrix algebra that has  $t\fdu{A}{B}$ generators that are decomposed with respect to $so(3,2)$. 
From the commutation relations of the $gl_{3+2}$ 
\begin{align}
    [t\fdu{A}{B}, t\fdu{C}{D}]=-\delta\fdu{A}{D}t\fdu{C}{B} +\delta\fdu{C}{B} t\fdu{A}{D}\,. \label{alg2}
\end{align} 
one can read out the irreducible generators in the $so(3,2)$ base $R$, $S_{AB}=S_{BA}$ and $T_{AB}=-T_{BA}$,
 as 
\begin{align}
R=t\fdu{C}{C}\,, && S_{AB}&=t_{A|B}+t_{B|A}-\frac{2}{d+2}t\fdu{C}{C}\eta_{AB}\,,  &&  T_{AB}=t_{A|B}-t_{B|A}\,. \label{eq15}
\end{align}
They close the commutation relations
\besubeqs
\begin{align}
    [T_{AB},T_{CD}&]= \eta_{BC} T_{AD}-\eta_{AC} T_{BD}-\eta_{BD} T_{AC}+\eta_{AD} T_{BC}\,,\\
    [T_{AB},S_{CD}&]= \eta_{BC}S_{AD}-\eta_{AC}S_{BD}+\eta_{BD}S_{AC}-\eta_{AD}S_{BC}\,,\\
    [S_{AB},S_{CD}&]= \eta_{BC}T_{AD}+\eta_{AC}T_{BD}+\eta_{BD}T_{AC}+\eta_{AD}T_{BC}\,.
\end{align}
\esubeqs
Here $R$ is associated with $1$ in $gl(V)$ which make it commute with everything, and allows us to truncate it away. This leads to a theory with a the conformal graviton $T_{AB}$ and a field $S_{AB} $ which is similar to the spin-three partially-massless field:
\begin{align}
    \omega=& \omega^{A,B} T_{AB} +\omega^{AB} S_{AB}.
\end{align}
Conformal graviton $\omega^{A,B}$ is the same one that was described in conformal gravity as a gauge theory in three dimensions, in \cite{Horne:1988jf}.
To obtain conformal gravity as a gauge theory of $so(3,2)$ one has to first fix the commutation relations of the conformal algebra $so(3,2)$ 
    \besubeqs\label{LPDK}
\begin{align}
[D,P^a]=&-P^a\,, & [J^{ab},P^c]&=P^a\eta^{bc}-P^b\eta^{ac}\,,\\
[D,K^a]=&K^a\,,  & [J^{ab},K^c]&=K^a\eta^{bc}-K^b\eta^{ac}\,,\\
[P^a,K^b]=&-J^{ab}+\eta^{ab}D\,, &[J^{ab},J^{cd}]&=J^{ad}\eta^{bc}-J^{ac}\eta^{bd}-J^{bd}\eta^{ac}+J^{bc}\eta^{ad},,\label{confalg}
\end{align}
\esubeqs
 (for $a=0,1,2$) and $P^a$ translations, $J^{ab}$ Lorentz boosts, $K^a$ special conformal tranformations,  and $D$ dilatation. 
Then it is required to evaluate the connection $\omega$ and the corresponding gauge parameter in the $so(3,2)$ 
\begin{align}
    \omega= \frac12 \varpi^{a,b} J_{ab}+e^aP_a +f^a K_a+ bD.
\end{align}
 The gauge is then fixed, and solving the equations of motion gives conformal gravity as a gauge theory of the $so(3,2)$ algebra. In other words, one can write the Chern-Simons action in the same way as in (2.1) however in latter case $\overline{\omega}$ is evaluated in the group of Lorentz rotations. 
Here, we follow that general procedure, for the non-relativistic limit of our conformal algebras.


First we linearize a theory over the Minkowski vacuum. 
This can be done by choosing $\omega_0=h^aP_a$ with $h^a=h_{\mu}{}^adx^{\mu}$, and the background metric $h_{\mu}{}^a=\delta_{\mu}{}^a$.
One can write the linearized equations and the linearized gauge symmetries as
\begin{align}
d\omega+\omega_0\wedge\omega+\omega\wedge\omega_0=0\,, && \delta\omega =d\xi+[\omega_0,\xi].\label{lineq}
\end{align}
The background field is $\omega_0$, and $\omega$  is the one-form evaluated in the Lie algebra. This algebra is one of the algebras of the so(3,2) irreducible modules that we obtain from the decomposition of the $\mathfrak{hs}(V)$, and 
by taking the non-relativistic limit of the algebra.
When we insert the field and gauge parameter expansion in the equations (\ref{lineq}) we  obtain linearized equations of motion and linearized gauge transformations 
for the  $t^{\Lambda}$ generators, field $\omega=\omega^{\Lambda}t_{\Lambda}$ and gauge parameter $\xi=\xi^{\Lambda}t_{\Lambda}$. That  will lead to 
\begin{align}
d\omega^{\Lambda}t_{\Lambda}+h^a\wedge\omega^{\Lambda}[P_a,t_{\Lambda}]=0, && \delta \omega^{\Lambda}t_{\Lambda}=d\xi^{\Lambda}t_{\Lambda}+h^a\wedge\xi^{\Lambda}[P_a,t_{\Lambda}].
\end{align}
 $\Lambda$  are indices in the light cone coordinates $\Lambda=\{a,+,-\}$, $\eta^{AB}$ is corresponding metric with $\eta_{+-}=\eta_{-+}=1$, and $t^{\Lambda}$ are generators of so(3,2) irreducible modules $T_{AB}$, $S_{AB}$ etc.. 


\subsection{In\"on\"u Wigner contraction }

To consider the Galilean limit we use In\"on\"u Wigner (IW) contraction 
\cite{Inonu:1953sp} of the algebra (2.7).
 We decompose an algebra $\mathfrak{g}$  in $\mathfrak{g}=\mathfrak{h}+\mathfrak{i}$ direct sum of vector spaces. For $\mathfrak{h}$ a subalgebra and $\mathfrak{i}$ an ideal. The generators of the ideal are rescaled with contraction parameter $\epsilon$, where  $\mathfrak{i}\rightarrow\epsilon\mathfrak{i}$. That gives for commutation relations
\begin{align}
    [\mathfrak{h},\mathfrak{h}]=\mathfrak{h},  &&  [\mathfrak{h},\mathfrak{i}]=\frac{1}{\epsilon}\mathfrak{h}+\mathfrak{i}, && [\mathfrak{i},\mathfrak{i}]=\frac{1}{\epsilon^2}\mathfrak{h}+\frac{1}{\epsilon}\mathfrak{i}.
\end{align}
If we take
$\epsilon\to\infty$, we need to obtain well defined limit. 
For this limit to hold, one needs to get $\mathfrak{h}$ as a subalgebra of $\mathfrak{g}$. If we had $[\mathfrak{h},\mathfrak{h}]=\mathfrak{h}+\epsilon \mathfrak{i}$ we would not have well defined limit $\epsilon\to\infty$. For the spin-2 case we are able to compare the GCA with the literature \cite{Bagchi:2009ca} and see that  this procedure leads us to the same algebra.

\section{Spin-2}

The first algebra we consider is the algebra described by the generator $T_{A,B}=-T_{B,A}$, the Young module $\YoungpAA$, and it was given in (2.7). We have the coordinates $t,x_i$ for $i=1,2$ and we refer to the coordinate t as the 0-th coordinate. 
The rescaled generators in the non-relativistic limit are those that are along the $x_i$ coordinates. If we rename the generators 
 of $so(3,2)$ algebra such that $T_{i+}=P_i,T_{0+}=H,T_{i-}=K_i,T_{0-}=K_0,T_{ij}=J_{ij},T_{0i}=B_i$, and $T_{+-}=-D$ we can decompose an algebra into 
\begin{align}
\mathfrak{i}=\{P_a,K_a,B_a\}, && \mathfrak{h}=\{J_{ab},D,H,K_0\}.
    \end{align}
When we take into account this decomposition, (2.7) and (2.11) we obtain GCA
\besubeqs
\begin{align}
    [D,P_a]&=-P_a && [D,H]=-H && [D,K_a]=K_a \\
    [D,K_0]&=K_0 &&[P_a,K_0]=B_{a} && [H,K_b]=-B_b 
     \\ [H,K_0]&=\eta_{00}D && [B_{b},H]=-\eta_{00}P_b && [B_{b},K_0]=-\eta_{00}K_b \\
    [J_{ab},P_c]&=P_a\eta_{bc}-P_b\eta_{ac} &&  [J_{ab},K_c]=K_a\eta_{bc}-K_b\eta_{ac}  &&  [J_{ab},B_{d}]=B_a\eta_{bd}-B_b\eta_{ad}
    \end{align}
    \esubeqs
\begin{align}
    [J_{ab},J_{cd}]=J_{ad}\eta_{bc}-J_{ac}\eta_{bd}-J_{bd}\eta_{ac}+J_{bc}\eta_{ad} \end{align}
in which we wrote only non-vanishing commutation relations. We write the field decomposed in the generators (3.1) of the algebra (3.2) and (3.3) 
\begin{align}
    \omega&= e^a P_a+\tau H +\tfrac12 \omega^{a,b}J_{ab}+\beta^{a}B_{a}-b D+f^{a}K_{a}+\kappa K_{0}\label{connection}
\end{align}
as well as the gauge parameter $\xi$
\begin{align}
    \xi&=\xi^{a+}P_a+ \xi^{0+}H +\tfrac12 \xi^{a,b}J_{ab}+\xi^{0a}B_a- \xi^{+-}D+\xi^{a-}K_a+ \xi^{0-}K_0.\label{gaugepar}
\end{align}
Inserting the gauge field $\omega$ and gauge parameter $\xi$ in the linearized equations (2.10) gives 
\begin{footnotesize}
\besubeqs
\begin{align}
    D:&&  -db-h^0\wedge\kappa=0 &&  \delta b&=d\xi^{+-}+h^0 \xi^{0-} \\
    P_i:&& de^i-h^j\wedge\omega^i{}_j-h^i\wedge b-h^0\wedge\beta^i=0 && \delta e^i&=d\xi^{i+}-h^j\xi^i{}_j-h^i\xi^{+-}-h^0\xi^{0i} \\
    H: && d\tau-h^0\wedge b=0 && \delta\tau&=d\xi^{0+}-h^0\xi^{+-} \\
    J_{ij}: && d\omega^{ij}J_{ij}=0 && \delta\omega^{ij}&=d\xi^{ij} \\
    B_j:&& d\beta^i-h^0\wedge f^i+h^i\wedge\kappa=0 && \delta\beta^i&=d\xi^{0i}-h^0\wedge\xi^{i-}+h^i\wedge\xi^{0-}\\
    K_i: && df^i=0 && \delta f^i&=d\xi^{i-} \\
    K_0: && d\kappa=0 && \delta\kappa&=d\xi^{0-}
          \end{align}
\esubeqs
\end{footnotesize}
Considering relations (1.14a),  the gauge transformation condition  for m and for 0 components
leads to $\delta b_m=\partial_m\xi^{+-}$, a gauge transformation for the field $b_m$ since $h_m{}^0=0$. And
we can fix a field $b_0$ to be zero using the parameter $\xi^{0-}=-\partial_0\xi^{+-}$.
From the equation (1.14a) 
 we consider mn and m0 components respectively
\besubeqs
\begin{align}
  -\partial_{m}b_{n}+\partial_{n}b_{m}=0 \\
 \partial_{0}b_{m}+\kappa_{m}=0.    \end{align}
\esubeqs
and obtain condition on the $b_m$ field, and definition of the field $\kappa_m$ in terms of the field $b_m$.\\
The following relations we consider are (1.14b). The gauge condition 
 we write for the m and 0 components. 
The m component gauge transformation consists of symmetric part with trace and antisymmetric part. The symmetric part gives
$ \delta e_{(mi)}=\frac{1}{2}(\partial_{m}\xi_i{}^++\partial_{i}\xi_m{}^+)-h_{mi}\xi^{+-}.$
 If we choose to fix the gauge parameter $\xi^{+-}$, we can set the trace of the $e_{mi}$ field to zero $e_m{}^m=0$ when $\xi^{+-}=\frac{1}{2}\partial_m\xi^{m+}$.
This gives for linearized gauge transformation of the field $e_{(mi)}$
\begin{align}
 \delta e_{(mi)}&=\frac{1}{2}(\partial_{m}\xi_i{}^++\partial_{i}\xi_m{}^+)-\frac{1}{2}h_{mi}\partial_l\xi^{l+}.
   \end{align}
The antisymmetric part 
$\delta e_{[mi]}$
can be fixed to zero by
$\xi_{im}=\frac{1}{2}(\partial_{m}\xi_i{}^+-\partial_{i}\xi_m{}^+).
$ To remember $e_{mi}$ is symmetric traceless field we define $e_{mi}\equiv\phi_{mi}$.
The 0 component of gauge transformation allows us to fix $e_{0}{}^i=0$ using the gauge fixing $\xi^{0i}=\partial_0\xi^{i+}$. 

Equation of motion (3.6b) 
 inserting the gauge fixings $e_{mi}=\phi_{mi}$, $e_{0}{}^i=0$ and $b_0=0$ is in components
\besubeqs
\begin{align}
mn: && \partial_{m}\phi_{ni}-\partial_{n}\phi_{mi}-\omega_{nim}+\omega_{min}-h_{mi}b_{n}+h_{ni}b_{m}=0 \\
m0: && -\partial_{0}\phi_{mi}-\omega_{0im}+\beta_{mi}=0.
\end{align}
\esubeqs
First of these equations, (1.26a) has symmetric, antisymmetric and a hook part. If we consider totally symmetric part we obtain the equality 0=0, which is reasonable because we know that the field $\omega_{min}$ is antisymmetric in the last two indices while entire equation already has antisymmetry in m and n. Antisymmetric part of the equation gives us  
\begin{align}\omega^{imn}-\omega^{\min}+\omega^{nim}=0.\label{eq1}\end{align} The hook part of the equation 
can using the equation (\ref{eq1}), and the property $\omega_{min}=-\omega_{mni}$ let us express 
\begin{align}
    \omega^{nim}=\frac{1}{2}(b^nh^{im}-b^mh^{in}-\partial^m\phi^{in}+\partial^n\phi^{im}).
\end{align}
The equation (3.9b) can be split in symmetric and antisymmetric part. Symmetric part is going to give \begin{align}
\beta_{(mi)}=\partial_{0}\phi_{(mi)}.
\end{align}
Since $e_{(mi)}$ is traceless we notice that $\beta_{m}{}^{m}$ is zero as well.
The antisymmetric part is going to define the antisymmetric part of $\beta_{mi}$ in terms of $\omega_{0im}$ such that $\beta_{[mi]}=\omega_{0im}.$ In the instance of conformal Galilean algebra, as it is already known there is not enough constraints to determine spin connection in terms of the remaining fields.

From the gauge transformation (3.6c) we obtain that from m component $\delta\tau_m=\partial_m\xi^{0+}$, while the parameter that is in 0  component we fixed already. That means $\delta\tau_0=\partial_0\xi^{0+}-\frac{1}{2}\partial_m\xi^{m+}$.
The equation of motion gives in components  
$   \partial_{m}\tau_{n}-\partial_{n}\tau_{m}=0$ and $
  \partial_{m}\tau_{0}-\partial_{0}\tau_{m}+b_{m}=0$
and they allow us to define $b_m$ as $b_{m}=\partial_0\tau_m-\partial_m\tau_0$. Here $\tau_0$ remains as undetermined component. 

The gauge condition with $J_{ab}$ generator leads for the components m and 0 to 
$\delta\omega_{mab}=\partial_{m}\xi_{ab}$ and $     \delta\omega_{0ab}=\partial_0\xi_{ab}$ 
respectively. The gauge condition for m component  has symmetric part which is equal to zero, antisymmetric part $\delta\omega^A_{mij}=\tfrac{1}{3} (\partial_{i}\xi_{jm} -  \partial_{j}\xi_{im} + \partial_{m}\xi_{ij})=0$ which is zero upon inserting the gauge fixing for $\xi_{ij}$, and a hook part $\delta\omega^H_{mij}=\tfrac{1}{3} (-2 \partial_{i}\xi_{jm} -  \partial_{j}\xi_{im} + \partial_{m}\xi_{ij})$.  The gauge transformation for the 0 component has only antisymmetric part by construction $\delta\omega_{0ab}=\tfrac{1}{2}(\partial_0\xi_{ab}-\partial_0\xi_{ba})$.
The equation from $J_{ab}$  in the mn components 
 has symmetric, hook and Riemann parts which are equally 0 because of $\omega_{mab}$ is antisymmetric in the last two indices, while the antisymmetric part of the equation is satisfied when the value for $\omega_{mab}$ (obtained from expressions with $P_i$) is inserted. The m0 components give the symmetric and hook equations which give 0=0 and antisymmetric part which defines equation 
\begin{align}
\partial_a\omega_{0bm}-\partial_b\omega_{0am}+\partial_m\omega_{0ab}=0    
\end{align}
for $\beta_{ab}$ taking into account that $\omega_{0ab}=\beta_{[ab]}$.

For $B_j$ we obtain the gauge transformations 
\begin{align}
m: &&    \delta\beta_{mi}&=\partial_m\xi^{0}{}_i+h_{mi}\xi^{0-} \\
0: &&   \delta\beta_{0i}&=\partial_0\xi^{0}{}_i-\xi_i{}^{-}
\end{align}
The first defines symmetric $\delta\beta^S_{mi}=\tfrac{1}{2}(\partial_m\xi^{0}{}_i+\partial_i\xi^{0}{}_m)+h_{mi}\xi^{0-} $ and antisymmetric part $\delta\beta^A_{mi}=\tfrac{1}{2}(\partial_m\xi^{0}{}_i-\partial_i\xi^{0}{}_m)$ of the gauge transformation for the field $\beta_{mi}$.
Inserting the value for the $\xi^{0-}$ and $\xi^{0i}$ we obtain the gauge transformation for the symmetric traceless part of the field $\delta\beta^S_{(mi)}=\tfrac{1}{2}(\partial_m\partial_0\xi_i{}^++\partial_i\partial_0\xi_{m}{}^+)-\tfrac{1}{2}h_{mi}\partial_0\partial_l\xi^{l+}$ field. We can also verify that the trace of the field $\beta_{mi}$ vanishes. The 0 component allows to fix gauge parameter $\xi^{i-}=\partial_0\xi^{0i}$ and set $\beta_0^i=0$. 
The equations with the $B_i$ generator can for mn components lead to the symmetric part and hook part that lead to 0=0 while the antisymmetric part of the equation gives the equation for the $\beta_{mn}$ field which is equal to the equation (3.13).

The m0 component of the equation will fix the symmetric, antisymmetric parts of the field $f_{ma}$ and the trace. The symmetric part of the field is 
\begin{align}
    f^S_{(ma)}=\partial_0\beta_{(ma)}-h_{ma}\kappa_0=\partial_0\partial_0\phi_{(ma)}-h_{ma}\kappa_0
\end{align}
while the antisymmetric part is defined by the antisymmetric part of the $\beta_{ij}$ field 
   $ f_{[ma]}^A=\partial_0\beta_{[ma]}.
$

The following gauge generator that we consider, $K_a$ gives gauge transformations 
of the symmetric traceless and antisymmetric part of $f_{mn}$ field such that 
$\delta f_{(ma)}=\tfrac{1}{2}\partial_m\xi_a{}^-+\partial_a\xi_m{}^--\frac{1}{2}h_{ma}\partial_l\xi^{l-}$.
While the antisymmetric part reads 
   $ \delta f_{[ma]}=\tfrac{1}{2}(\partial_m\partial_0\partial_0\xi_a{}^+-\partial_a\partial_0\partial_0\xi_m{}^+).
$
The mn equation consists of the symmetric and hook parts which give 0=0, and the antisymmetric part which gives the same equation as (3.13).
The m0 equation consists of the symmetric and antisymmetric parts. Symmetric part gives equation that relates $f_{0i}$ and $\kappa_0$ \begin{align}
    \frac{1}{2}(\partial_mf_{0a}+\partial_af_{0a})+h_{ma}\partial_0\kappa_0=0
\end{align}
while the antisymmetric part gives the equation that relates $f_{0a}$ and the antisymmetric part of the $\beta_{ij}$ field
\begin{align}
    \frac{1}{2}(\partial_mf_{0a}-\partial_af_{0m})-\partial_0\partial_0\beta_{ma}=0.
\end{align}
The gauge transformations with the following generator $K_0$ do not have any more gauge parameter that we can fix, so they only define the transformation rules of the fields, and give equation for $\tau_m$ and $\tau_0$ fields. 

After we have fixed the gauge, we notice that the undetermined fields that remain are 
$\tau_0,\phi_{mi},\tau_m,\beta_{[mi]},\kappa_0,f_{0i}$. We summarize the fields in the Table 1.
\begin{table}[ht!]
\begin{center}
\begin{tabular}{|c|l|l|c|} \hline
Generator & Field \\ \hline
D& $b_0=0$     \\
&$ b_m=\partial_0\tau_m-\partial_m\tau_0$    \\ \hline
$P_i $ & $e_{(mi)}\equiv\phi_{mi},e_{[mi]}=0,e_m{}^m=0$   \\
& $e_{0i}=0$  \\ \hline
H& $\tau_m$    \\
& $\tau_0$   \\ \hline
$J_{ij}$& $\omega^{nim}=\frac{1}{2}(b^nh^{im}-b^mh^{in}-\partial^m\phi^{in}+\partial^n\phi^{im})$  \\
& $\omega_{0im}$   \\ \hline
$B_j $ & $\beta_{(mi)}=\partial_0\phi_{mi},\beta_{[mi]}=\omega_{0im}$  \\
&$\beta_{0i}=0$  \\
 \hline
$K_i$ & $f_{(mi)}=\partial_0\partial_0\phi_{mi}-h_{mi}\kappa_0$  \\
& $f_{[mi]}=\partial_0\beta_{[mi]}$   \\
 & $f_{0i}$   \\ \hline
$K_0$& $\kappa_0$  \\
&$ \kappa_n=\partial_0 b_m$  \\ \hline
 \end{tabular}
 \caption{Spectrum of fields in Galilean conformal gravity in 3d.}
\end{center}
 \end{table}

The gauge transformations of the fields are listed in the Appendix in Table 3. One has to notice that $\beta_{[mi]}$ is equal to 0 component of the spin connection which shows that due to degeneracy of the metric in this case, we cannot determine entire spin connection in terms of the other fields, as it is known for Galilean conformal gravity. One can speculate about adding the central extension, however GCA does not admit the central extension as Galilean algebra. However it does admit different kind of central extension. There is also semi-GCA, which in 3d corresponds to $BMS_4$ algebra \cite{Bagchi:2009my}. Analysis of the $BMS_4$ was done in \cite{Barnich:2022bni} and it would be interesting to see relation from the side of semi-GCA in 3d. Here, we first want to see the contribution from the spin-3 field. 

\section{Spin-3}

The algebra that describes spin-3 field is defined by the $S_{AB}=S_{BA}$ generator, and a Young module $\YoungpB$. We define the names of the generators in the analogous way as in the previous scenario, with a difference that here we add a prefix S- to the generators and s- to the  fields that appeared es well above. This way we additionally point out that they are defined in relation with $S_{AB}$ generator. We keep the notation for the field $\omega_{\mu ab}$.  The generators that appear here and do not have antisymmetric analog in spin-2 case are $t_{++}$, $t_{--}$, $t_{00}\equiv SC$ and $SJ_{kj}$ for when $j=k$.  We decompose the generators analogously as above by setting the generators with space components in the ideal $\mathfrak{i}$, and the remaining components in $\mathfrak{h}$.
\begin{align}
    \mathfrak{i}=\{ SP_a,SK_a,SB_a \}&&  \mathfrak{h}&=\{ SH,SK_0,t_{++},t_{--},SJ_{jk},SC \} \label{eq41}
\end{align}
 To achieve the closure of the algebra $\mathfrak{h}$ we set the trace components of the $SJ_{ij}$ in $\mathfrak{h}$ as well. One can notice that due to the trace condition there is no generator $SD$ appearing, but only generators $SJ_{ii}$ for $i=0,1,2$. The Lie algebra valued one and zero forms after decomposition in the generators (\ref{eq41}) read 
{\footnotesize      
\begin{align}
    \omega&= se^{a}SP_{a}+s\tau SH +\tfrac12 \omega^{ab}SJ_{ab}+s\beta^{a}SB_{a}+\gamma SC+\frac12\omega^{++}t_{++}+\frac12\omega^{--}t_{--}+sf^{a}SK_{a}+s\kappa SK_{0}\,,\\
    \xi&= \xi_{a+}SP_a+ \xi^{0+}SH +\tfrac12 \xi^{ab}SJ_{ab}+\xi^{0a}SB_a+ \xi^{00}SC+\frac12\xi^{++}t_{++}+\frac12\xi^{--}t_{--}+\xi^{a-}SK_a+\xi^{0-}SK_0 \,.
\end{align} } 
From (2.5) one can read out the commutators of the spin-3 generators and $P_{\mu}$ 
\begin{align}
    [P_i, t_{j+}]&=-\eta_{ij}t_{++}\,, & [P_i, t_{j-}]&=t_{ij}+\frac{1}{2}t\fdu{m}{m}\eta_{ij}\,, & [P_i, t_{--}]&=2t_{i-}\,, \\ [P_i, t_{++}]&=0\,, &
    [P_i, t_{jk}]&=-\eta_{ij}t_{k+}-\eta_{ik}t_{j+}\,. \label{algp3}
\end{align}
and insert the decomposition (\ref{eq41}) and take IW contraction (2.11). That will give non-vanishing commutation relations
\besubeqs
\begin{align}
     [H,SK_b]&=SB_b && [H,SC]=-2\eta_{00}SH && [H,t_{--}]=2SK_0 \\
     [P_a,SJ_{cd}]&=-\eta_{ac}SP_{d}-\eta_{ad}SP_c && [P_a,t_{--}]=2SK_a &&  [P_a,SK_0]=SB_a\\
    [H,SK_0]&=SC+\frac{1}{2}(SJ_a^a+SC)\eta_{00} && [H,SH]=-\eta_{00}t_{++} && [H,SB_d]=-\eta_{00}SP_d. 
\end{align}
\esubeqs
When we insert these commutation relations in the linearized gauge equations and gauge transformations (2.10) we can write the equations 
\besubeqs
\begin{align}
SP_a: &&   dse^a-h^l\wedge\omega^a{}_l+h^0\wedge s\beta^a=0 && \delta se^a&=d\xi^{a+}-h^l\wedge\xi^a{}_l+h^0\wedge\xi^{a0}\\
SH: && ds\tau+2h^0\wedge\gamma=0 && \delta s\tau&=d\xi^{0+}+2h^0\wedge\xi^{00}\\
SJ^{ab}: &&  d\omega^{ab}-h^0\wedge s\kappa h^{ab}=0 
&&  \delta \omega^{ab}&=d\xi^{ab}-h^0\wedge\xi^{0-}\eta^{ab}  \\ 
SB^a: && ds\beta^a+h^a\wedge s\kappa+h^0\wedge sf^a=0 && \delta s\beta^a&=d\xi^{a0}+h^a\wedge\xi^{0-}+h^0\wedge\xi^{a-} \\
SC: && d\gamma+\frac{1}{2}h^0\wedge s\kappa=0 && \delta\gamma&=d\xi^{00}+\frac{1}{2}h^0\wedge\xi^{0-} \\
t_{++}: && \frac{1}{2}d\omega^{++}+h^0\wedge s\tau=0 && \frac{1}{2}\delta\omega^{++}&=\frac{1}{2}d\xi^{++}+h^0\wedge\xi^{0+} \\
t^{--}: && d\omega^{--}t_{--}=0 && \delta\omega^{--}&=d\xi^{--} \\
SK_a: && dsf^a+h^a\wedge\omega^{--}=0 && \delta sf^a&=d\xi^{a-}+h^a\wedge\xi^{--}\\
SK_0: && ds\kappa+h^0\wedge\omega^{--}=0 && \delta s\kappa&=d \xi^{0-}+h^0\wedge\xi^{--}
\end{align} \label{eq47}
\esubeqs
To solve the system of equations (\ref{eq47})
we start from $t^{++}$  gauge transformations 
\begin{align}
m: &&    \delta\omega_{m}{}^{++}&=\partial_m\xi^{++} \\
0: && \delta\omega_0{}^{++}&=\partial_0\xi^{++}+2\xi^{0+}
\end{align}
where in the second equation we can fix $\omega_0{}^{++}=0$ with $\xi^{0+}=-\frac{1}{2}\partial_0\xi^{++}$.
The corresponding equation in components 
leads to equation $\partial_m\omega_n{}^{++}=\partial_n\omega_m{}^{++}$ for mn components, and for m0 components to $s\tau_m=-\frac{1}{2}\partial_0\omega_m{}^{++}$. That expresses  the $s\tau_m$ field in terms of $\omega_m^{++}$.

The gauge transformations with the SH generator lead to $\delta s\tau_m=\partial_m\xi^{0+}$ for the m component, and allow for the fixing $s\tau_0=0$ when $\xi^{00}=\tfrac{1}{2}\partial_0\xi^{0+}$. The corresponding equation then gives $\partial_m s\tau_n=\partial_n s\tau_m$ and $\gamma_m=\frac{1}{2}\partial_0 s\tau_m$. Where the latter equation expresses the field $\gamma_m$ in terms of the $s\tau_m$ field, which we already expressed in terms of $\omega_m^{++}$. $s\tau_m$ reads $s\tau_m=-\tfrac{1}{4}\partial_0\partial_0\omega_m^{++}$.
 So far we have one undetermined field, $\omega_m^{++}$.
 
 The gauge transformations with the $SP_a$ generator give the symmetric and antisymmetric part of the $se_{ma}$ field. The symmetric part can be gauged away so that $se_{(ma)}=0$ when $\xi_{dm}=\tfrac{1}{2}(\partial_m\xi_d^++\partial_d\xi_m^+)$. The antisymmetric part remains unfixed. While
the linearized gauge transformations of the field $se_{md}$ read $\delta se_{[md]}=\tfrac{1}{2}(\partial_m\xi_d^+-\partial_d\xi_m^+)$.
 
 The mn components of the equation have standardly symmetric, antisymmetric and a hook part. Symmetric part is equally 0, antisymmetric part reads
 \begin{align}
  \omega^{imn} -  \omega^{min} +  \omega^{nim} + \partial^{i}se^{mn}  -  \partial^{m}se^{in} + \partial^{n}se^{im} =0
 \end{align}
 here we took into account that $se_{im}$ is an antisymmetric field. When we take into account that $\omega^{imn}$ is symmetric in last two indices the equation reduces to equation for $se^{mn}$ field $\partial^{i}se^{mn}  -  \partial^{m}se^{in} + \partial^{n}se^{im} =0$. The hook part of the equation gives
 \begin{align}
    - \omega^{imn} - \omega^{min} + \partial^{i}se^{nm} + \partial^{m}se^{ni}=0 
 \end{align}
 taking that $se_{ma}$ is antisymmetric.
 Taking into account that $\omega^{imn}$ is symmetric in last two indices gives that symmetric part of the $\omega^{imn}$ field remains undetermined, while the hook part is defined by $\omega^{(H)imn}=-\frac{1}{3}(\partial^ise^{mn}+\partial^mse^{in})$. 

The component 0 of the gauge transformation allows to fix $\xi^{a0}=\partial_0\xi^{a+}$ and $se_0{}^a=0$. The equation of motion for the component m0 contains the gauge fixings and reads 
\begin{align}\partial_0 se_{ma}+\omega_{0am}+s\beta_{ma}=0.\label{eq2}\end{align} The equation consists of the symmetric and antisymmetric parts. Since $s\omega_{0am}$ is symmetric in last two indices and $se_{ma}$ is gauge fixed to be a antisymmetric field, equation defines $s\beta_{(ma)}=-\omega_{0am}$ and the antisymmetric part of the $s\beta_{[am]}=\partial_0se_{ma}$.  

The gauge transformation with the generator $SJ^{ab}$  reads $\delta \omega_{mab}=\partial_m\xi_{ab}$ and it has  symmetric $\delta\omega^{(mij)}=\tfrac{1}{3} (\partial^{i}\partial^{j}\xi^{+}{}^{m} + \partial^{i}\partial^{m}\xi^{+}{}^{j} + \partial^{m}\partial^{j}\xi^{+}{}^{i})$, antisymmetric $\delta\omega^{[mij]}=0$ and hook part $\delta\omega^{(H)mij}=\tfrac{1}{6} (-2 \partial^{j}\partial^{i}\xi^{+}{}^{m} + \partial^{m}\partial^{i}\xi^{+}{}^{j} + \partial^{m}\partial^{j}\xi^{+}{}^{i})$. The 0 component of the transformation gives $\delta \omega_{0}{}^{ab}=\partial_0\xi^{ab}-\xi^{0-}h_{ab}$. Using the $\xi^{0-}$ we can fix a trace of $\omega_0^{ab}$.
Equations of motion in the mn component give equation $\partial_m\omega_{nab}-\partial_n\omega_{mab}=0$ which can be divided in components to give equations for the field $\omega_{(mab)}$ and for the hook part of the field $\omega^{(H)}_{mab}$.  Symmetric, antisymmetric, and Riemann component are equally zero, while hook component reads
\begin{align}
    \partial^{i}\omega^{jmn} -  \partial^{j}\omega^{imn} -  \partial^{j}\omega^{min} -  \partial^{j}\omega^{nim} + \partial^{m}\omega^{jin} + \partial^{n}\omega^{jim}=0
\end{align}
The symmetric equation from m0 component of equations of motion gives
\begin{align}
   & \partial_{0}{} \omega^{abm} + \partial_{0}{} \omega^{bam} - 2 \partial_{0}{} \omega^{mab} + 2 sk^{m} h^{ab} -  sk^{b} h^{am} -  sk^{a} h^{bm} -  \\ \nonumber &\partial^{a}\omega_{0}{}^{bm} -  \partial^{b}\omega_{0}{}^{am} + 2 \partial^{m}\omega_{0}{}^{ab}=0
\end{align}
antisymmetric part is 0, and the hook part leads to \begin{align}
   & \partial_{0}{} \omega^{abm} + \partial_{0}{} \omega^{bam} - 2 \partial_{0}{} \omega^{mab} + 2 sk^{m} h^{ab} -  sk^{b} h^{am} -  sk^{a} h^{bm} - \\ \nonumber & \partial^{a}\omega_{0}{}^{bm} -  \partial^{b}\omega_{0}{}^{am} + 2 \partial^{m}\omega_{0}{}^{ab}=0.
\end{align}
The combination of these equations gives that
$    sk^{m}=   \partial_{0}{} \omega^{am}{}_{a} - \partial_{a}\omega_{0}{}^{ma}.
$

The gauge transformation with $SB^a$ generator defines $\delta s\beta_{(ma)}=\frac{1}{2}(\partial_m\xi^0{}_a+\partial_a\xi^{0}{}_m)+h_{ma}\xi^{0-}$ and $\delta s\beta_{[ma]}=\frac{1}{2}(\partial_m\xi^0{}_a-\partial_a\xi^{0}{}_m)$ for m component. The trace of the $s\beta_{ma}$ is also vanishing since trace of $\omega_{0am}$ is gauge fixed to zero with $\xi^{0-}$. The 0 component allows to fix $s\beta_0{}^a=0$ when $\xi^{a-}=-\partial_0\xi^{0a}$.
The equations, for mn components, relate $s\kappa_a$ and $s\beta_{ab}$.
The symmetric part of the equation is equally zero, while the antisymmetric part gives equation for the antisymmetric part of $s\beta^{ab}$
\begin{align}
    \partial^{j}s\beta^{[mn]} -  \partial^{m}s\beta^{[jn]} + \partial^{n}s\beta^{[jm]}=0.
\end{align}
The hook component
\begin{align}
    \tfrac{1}{3} (- s\kappa^{n} h^{jm} -  s\kappa^{m} h^{jn} + 2 s\kappa^{j} h^{mn} -  \partial^{j}s\beta^{mn} -  \partial^{j}s\beta^{nm} + \partial^{m}s\beta^{jn} + \partial^{n}s\beta^{jm})=0
\end{align} can be contracted with $h^{ab}$ to give $s\kappa^m$ in terms of $s\beta_{ab}$
\begin{align}\tfrac{1}{3} (- s\kappa^{m} -  \partial_{a}s\beta^{ma} + \partial^{m}s\beta^{a}{}_{a})=0.\end{align}
The m0 equation of $SB^a$ gives $s\kappa_0=\frac{1}{2}\partial_0s\beta_m{}^m=0$ taking into consideration that $s\beta_0{}^a$ is gauge fixed to zero, as well as $\omega_{0a}{}^a$, and that we will be able to fix a trace of $sf_a{}^a$ field. This means that one has $sf_{am}=\partial_0s\beta_{am}$. 

The SC gauge transformation gives the transformations for $\delta\gamma_m=\partial_m\xi^{00}$ and $\delta\gamma_0=\partial_0\xi^{00}+\frac{1}{2}\xi^{0-}$. The equations define $\partial_{[m}\gamma_{n]}=0$ and $s\kappa_m$ in terms of $\gamma_m$ and $\gamma_0$, such that $s\kappa_m=\partial_m\gamma_0-\partial_0\gamma_m$.

The transformation with $SK_a$ defines 
$\delta sf_{(ma)}=\frac{1}{2}(\partial_m\xi_{a-}+\partial_a\xi_{m-})+h_{ma}\xi^{--}$ and $\delta sf_{[ma]}=\frac{1}{2}(\partial_m\xi_{a-}-\partial_a\xi_{m-})$. The component 0 defines $\delta sf_{0a}=\partial_0\xi^{a-}$.
The equations of motion in mn component, vanish for the symmetric part, for antisymmetric  they give equation for antisymmetric part of $sf_{[mn]}$ so that
$ \partial^{j}sf^{[mn]}  -  \partial^{m}sf^{[jn]}  + \partial^{n}sf^{[jm]} =0$. The hook component reads 
$\tfrac{1}{3} (- \omega^{--}{}^{n} h^{jm} -  \omega^{--}{}^{m} h^{jn} + 2 \omega^{--}{}^{j} h^{mn} -  \partial^{j}sf^{mn} -  \partial^{j}sf^{nm} + \partial^{m}sf^{jn} + \partial^{n}sf^{jm})=0$ and after contracting with $h^{jm}$ it defines $\omega^{n--}=\partial^nsf_a{}^a-\partial_asf^{na}$. 
The 0 component defines the field $\omega_0^{--}=-\frac{1}{2}\partial_m sf^m{}_{0}$.

For the $SK_0$ the transformation in the m component gives $\delta s\kappa_m=\partial_m\xi^{0-}$ and in 0 component one obtains $\delta s\kappa_0=\partial_0\xi^{0-}+\xi^{--}$. The equations of motion give $\partial_{[n}s\kappa_{m]}=0$ and $\omega_m^{--}=\partial_ms\kappa_0-\partial_0s\kappa_m$ for mn and m0 components respectively. 
The remaining fields are $se_{[am]}, \omega_m^{++},\omega^{(mab)}\equiv\phi^{mab},\omega_0{}^{ab},sf_{0i}$,  and $\gamma_0$. The summary of the fields is given in Table 2, while their gauge transformations are written in the Appendix, in Table 4.

\begin{table}[ht!]
\begin{center}
\begin{tabular}{|c|l|l|c|} \hline
Generator & Field \\ \hline
$t_{++}$& $\omega_0{}^{++}=0$      \\
&$ \omega_m{}^{++}$     \\ \hline
$SP_i $ & $se_{(mi)}=0$   \\
& $se_{[mi]}$  \\
& $se_{0i}=0$   \\ \hline
SH& $s\tau_m=-\frac{1}{2}\partial_0\partial_0\omega_m{}^{++}$   \\
& $s\tau_0=0$   \\ \hline
$SJ_{ij}$& $\omega^{(H)imn}=-\frac{1}{3}(\partial^ise^{mn}+\partial^mse^{in}),\omega^{[imn]}=0$  \\
&$\omega^{(imn)}\equiv\phi^{imn}$  \\
& $\omega_{0im}$   \\ \hline
$SC$ & $\gamma_0 $ \\ 
& $\gamma_m =-\frac{1}{4}\partial_0\partial_0\omega_m^{++}$ \\ \hline
$SB_j $ & $s\beta_{(mi)}=-\omega_{0im},s\beta_{[im]}=\partial_0se_{mi}$  \\
&$s\beta_{0i}=0$  \\
 \hline
$SK_i$ & $sf_{(mi)}=\partial_0s\beta_{im}$  \\
 & $sf_{0i}$   \\ \hline
$SK_0$& $s\kappa_0=0$  \\
&$ s\kappa^m=-\partial_as\beta^{ma}$  \\ \hline
$t_{--}$& $\omega_0{}^{--}=-\frac{1}{2}\partial_msf_{0}{}^m$  \\
&$ \omega_m{}^{++}=-\partial_0 s\kappa_m$  \\ \hline
 \end{tabular}
 \caption{Spectrum of fields in spin-3 Galilean conformal gravity.}
\end{center}
 \end{table}
It is clear  that this number of fields remains due to degeneracy of metric. If we compare the number of fields in this non-relativistic limit of the conformal spin-3 case and the number of remaining fields of conformal spin-3 in ultra-relativistic limit in \cite{Lovrekovic:2021dvi}, we can see that here there is more undetermined fields remaining. There we had $\phi_{mni}, \omega^{0++},s\beta_{(mi)},\gamma_m$ and $\gamma_0$. Here we have $se_{[am]}$  and $sf_{0m}$ that do not have analog in ultra-relativistic limit, while there we had $\omega_0^{++}$ which was here gauge fixed to zero. The other fields are either obtained from the same relativistic field, or are determined in terms of each other. If we wanted to compare the ultra-relativistic and non-relativistic limit of the spin-3 fields we would have to consider semi-GCA. In that case we would have similar situation as in the two dimensions when one can interchange the time and space component and relate these two limits.

\section{Conclusion}

We have considered Galilean limit of conformal algebra for spin-2 and spin-3 fields, and studied the gauge theory of these algebras. We obtained the spectrum of the theories. In spin-2 the remaining fields were
$\tau_0,\phi_{mi},\tau_m,\beta_{[mi]},\kappa_0,f_{0i}$. 
While in the spin-3 case these were $se_{[am]}, \omega_m^{++},\phi_{mab},\omega_{0ab},sf_{0i}$,  and $\gamma_0$. The action that is obtained in these theories does not include all the fields that appear in the spectrum. This is due to the degeneracy of the metric, and it is similar to the situation that happens for the non-conformal case \cite{Bergshoeff:2017btm}.
Interesting thing to consider would be to see if in the non-relativistic case of the gauged theory of the  s-GCA one can obtain the results that modify Newton-Cartan gravity in the analogous way that relativistic conformal gravity modifies Einstein gravity. One of the aspects in which  relativistic conformal gravity has been studied \cite{Mannheim:2011ds}  showed that one can obtain galactic rotation curves without the addition of the dark matter. In the non-conformal case, Galilean algebra had to be supplemented by the central extension, so that Bargmann algebra leads to the Newton-Cartan gravity when the theory is a gauge theory. When one tries to apply the analogous thinking to the conformal case, this is not possible, because GCA does not admit the same kind of central extension as Galilean algebra \cite{Bagchi:2009my}. The reason for this can be seen if one tries to calculate Jacobi identities for the generators ($B_i,P_i,D$) where $B_i$ generators are the In\"on\"u Wigner contracted generators of the Lorentz generators. One of the interesting things for the future research would be to study s-GCA or certain extension of the GCA to obtain the algebra gauging which gives the analogy of the non-relativistic conformal gravity. Using the same type of extension one could then implement to obtain also higher spin extension of the type \cite{Grigoriev:2019xmp} of GCA.

\section{Acknowledgements}
This work was supported by the Hertha Firnberg grant T 1269-N of the Austrian Science Fund FWF.



\section{Appendix}

Table of gauge transformations and fixed gauge parameters for the spin-2 case of GCA.
\begin{table}[ht!]
\begin{tabular}{|c|l|l|c|} \hline
D&$\delta b_0=0$     &   $\xi^{0-}=-\partial_0\xi^{+-}$\\
&$\delta b_m=\partial_m\partial_0\xi^{0+}   $  &   \\ \hline
$P_i $ & $\delta e_{[mi]}=0$  & $\xi_{i,m}=\frac{1}{2}(\partial_m\xi_i{}^+-\partial_i\xi_m{}^+)$  \\
& $\delta e_{(mi)}=\frac{1}{2}(\partial_m\xi_i{}^++\partial_i\xi_m{}^+)-\frac{1}{2}h_{mi}\partial_l\xi^{l+}$ & \\
& $\delta e_{0i}=0$ & $\xi_{0,i}=\partial_o\xi_i{}^+$ \\ \hline
H& $\delta\tau_m=\partial_m\xi^{0+} $  &  \\
& $\delta\tau_0=0$ & $\xi^{+-}=\partial_0\xi^{0+}$  \\ \hline
$J_{ij}$& $\delta\omega_m{}_{ij}=\frac{1}{2}\partial_m(\partial_j\xi_i{}^+-\partial_i\xi_j{}^+)-\frac{1}{4}h_m{}^i\partial_l(\partial_j\xi^{l+}-\partial^l\xi_j{}^+)$ & \\
& $\delta \omega_{0ij}=\frac{1}{2}\partial_0(\partial_j\xi_i{}^+-\partial_i\xi_j{}^+)$ &  \\ \hline
$B_j $ & $\delta\beta_{(mj)}=\frac{1}{2}(\partial_m\partial_0\xi_j{}^++\partial_j\partial_0\xi_m{}^+)-h_{mj}\partial_l\partial_0\xi^{l+}$ & \\
&$\delta\beta_{[mj]}=\frac{1}{2}(\partial_m\partial_0\xi_j{}^+-\partial_j\partial_0\xi_m{}^+)$ & \\
& $\delta\beta_0{}^j=0$ & $\xi^{j-}=-\partial_0\xi_0{}^j$ 
\\ \hline
$K_i$ & $\delta f_{(mi)}=\frac{1}{2}(-\partial_m\partial_0\partial_0\xi_i{}^+-\partial_i\partial_0\partial_0\xi_m{}^+)+\frac{1}{2}h_{mi}\partial_l\partial_0\partial_0\xi^{l+}$ & \\
& $\delta f_{[mi]}=\frac{1}{2}(-\partial_m\partial_0\partial_0\xi_i{}^++\partial_i\partial_0\partial_0\xi_m{}^+)$ &  \\
&$\delta f_m{}^m=-\partial_m\partial_0\partial_0\xi^{m+}$ & \\ 
& $\delta f_0{}^i=-\partial_0\partial_0\partial_0 \xi^{i+}$ &  \\ \hline
$K_0$& $\delta\kappa_{m}=-\partial_m\partial_0\partial_0\xi^{0+}$ & \\
&$ \delta\kappa_0=-\partial_0\partial_0\partial_0 \xi^{0+} $ & \\ \hline
 \end{tabular}
 \caption{Gauge transformations and gauge parameters for the spin-2 case.}
 \end{table}
 \newpage
Table of gauge transformations and fixed gauge parameters for the spin-3 case.
\begin{table}[ht!]
\begin{tabular}{|c|l|l|c|} \hline
$t^{++}$ & $\delta\omega_0^{++}=0$ & $\xi^{0+}=-\frac{1}{2}\partial_0\xi^{++}$  \\ 
& $\delta\omega_m^{++}=\partial_m\xi^{++}+2s\tau_m\xi^{0+}$& \\ \hline
$SP_i $ & $\delta se_{(mi)}=0$  & $\xi_{im}=\frac{1}{2}(\partial_m\xi_i{}^++\partial_i\xi_m{}^+)$  \\
& $\delta se_{[mi]}=\frac{1}{2}(\partial_m\xi_i{}^+-\partial_i\xi_m{}^+)-\frac{1}{2}h_{mi}\partial_l\xi^{l+}$ &  \\
& $\delta se_{0i}=0$ & $\xi_{0i}=\partial_0\xi_i{}^+$ \\ \hline
SH& $\delta s\tau_m=\partial_m\xi^{0+} $  &  \\
& $\delta s\tau_0=0$ & $\xi^{00}=\frac{1}{2}\partial_0\xi^{0+}$  \\ \hline
$SJ_{ij}$ & $\delta\omega_{(mij)}=\frac{1}{2}(\partial_i\partial_j\xi_{m+}+\partial_i\partial_m\xi_{j+}+\partial_m\partial_j\xi_{i+})$ &  \\
&$\delta \omega_{(H)mij}=\frac{1}{6}(-2\partial_i\partial_j\xi_{m+}+\partial_i\partial_m\xi_{j+}+\partial_m\partial_j\xi_{i+})$  & \\ 
& $\delta \omega_{[mij]}=0$ &  \\
& $\delta \omega_{0ab}=\partial_0\xi_{ab}-h_{ab}\xi^{0-}$ & $\xi^{0-}=-\partial_0\xi_a{}^a$  \\ \hline
$SB_j $ & $\delta s\beta_{(mj)}=\frac{1}{2}(\partial_m\partial_0\xi_j{}^++\partial_j\partial_0\xi_m{}^+)-h_{mj}\partial_l\partial_0\xi^{l+}$ & \\
&$\delta s\beta_{[mj]}=\frac{1}{2}(\partial_m\partial_0\xi_j{}^+-\partial_j\partial_0\xi_m{}^+)$ & \\
& $\delta s\beta_0{}^j=0$ & $\xi^{j-}=-\partial_0\xi_0{}^j$ 
\\ \hline
SC & $\delta \gamma_m=\partial_m\xi^{00}$ & \\
 &$\delta\gamma_0=\partial_0\xi^{00}+\frac{1}{2}\xi^{0-}$ &  \\ \hline
$SK_i$ & $\delta sf_{(mi)}=\frac{1}{2}(-\partial_m\partial_0\partial_0\xi_i{}^+-\partial_i\partial_0\partial_0\xi_m{}^+)+h_{ma}\xi^{--}$ & \\
& $\delta sf_{[mi]}=\frac{1}{2}(-\partial_m\partial_0\partial_0\xi_i{}^++\partial_i\partial_0\partial_0\xi_m{}^+)$ &  \\
&$\delta sf_m{}^m=0$ & $\xi^{--}=\frac{1}{2}\partial_m\partial_0\partial_0\xi^{m+}$ \\ 
& $\delta sf_0{}^i=-\partial_0\partial_0\partial_0 \xi^{i+}$ &  \\ \hline
$SK_0$& $\delta s\kappa_{m}=-\partial_m\partial_0\partial_l\xi^{l+}$ & \\
&$ \delta s\kappa_0=-\partial_0\partial_0\partial_l \xi^{l+} +\frac{1}{2}\partial_m\partial_0\partial_0\xi^{m+}$ & \\ \hline
 \end{tabular}
 \caption{Gauge transformations and gauge parameters for the spin-3 case.}
 \end{table}

\footnotesize
\bibliographystyle{templatebib}
\bibliography{megabib.bib}

\end{document}